\documentstyle[twocolumn,prl,aps,epsfig]{revtex}
\def\vec#1{{\bf #1}}
\begin{document}
%%%%%
\title{New Lower Bound on Fermion Binding Energies}
\author{Olivier {\sc Juillet} and Sonia {\sc  Fleck}}
\address{Institut de Physique Nucl\'eaire,
Universit\'e Claude Bernard--CNRS-IN2P3\\
43, boulevard du 11 Novembre 1918,
F 69622 Villeurbanne, France}
\author{Lukas {\sc Theu\ss l} and Jean-Marc {\sc Richard}}
\address{ Institut des Sciences Nucl\'eaires,
Universit\'e Joseph Fourier--CNRS-IN2P3\\
53, avenue des Martyrs,
F 38026 Grenoble, France}
\author{K{\'a}lm{\'a}n {\sc Varga}}
\address{Solid State Division, Oak Ridge National Laboratory,
Oak Ridge, Te 37381, USA\\
and Institute for Nuclear Research of the Hungarian Academy of 
Sciences, Debrecen, Hungary}
\maketitle
\begin{abstract}
We derive a new lower bound for the ground state energy 
$E^{\rm F}(N,S)$ of $N$ fermions with total spin $S$ in terms of binding 
energies $E^{\rm F}(N-1,S\pm 1/2)$ of $(N-1)$ fermions. Numerical 
examples are provided for some simple short-range or confining potentials.
\end{abstract}
There is a persisting interest in deriving lower bounds on the
binding energy of $N$-particle systems.  Some early investigations
were motivated by fundamental studies on the thermodynamic limit or
the stability of matter \cite{FIRU66}.  Also, the search for a lower
bound is rather natural once an upper bound is provided by variational
estimates.
 
Many efforts have been devoted in particular to express a
bound on the binding energy $E_N$ of a $N$-body system in terms of a
$(N-1)$-body energy $E_{N-1}$ with modified constituent mass or
interaction strength.  Thanks to a proper account for the
center-of-mass motion of the subsystems \cite{HAPO67}, the situation
is now rather satisfactory in the zero-spin boson sector: the
bound of Hall and Post \cite{HAPO67} is saturated for the harmonic oscillator and approaches closely
the exact result for many other potentials.  An extension has been
derived for three and four particles with different masses
\cite{BAMA93} where, again,  saturation is  obtained in the case of harmonic
forces.

The situation is far more difficult for fermions.  The elegant bound
derived by L\'evy-Leblond \cite{FIRU66} suffers from the fact that the
energy of $(N-1)$-body subsystems is replaced by its rest energy
without including its overall kinetic energy.  As a result, saturation
is never reached.

A significant but partial improvement was obtained  by
Basdevant and Martin \cite{BAMA96}, who used subtle convexity
inequalities.  Their bound improves that of L\'evy-Leblond in some
cases, and  becomes exact for the harmonic oscillator. 
Their approach, however, is restricted to confining potentials $r^q$,
$q\ge 1$, or superpositions of such power-law potentials with positive
weight factors.

In this letter, we use group-theoretical techniques to derive a
general lower bound on the $N$-body energy of an interacting system of
particles with internal degrees of freedom.  The decomposition of the
Hamiltonian is supplemented by considerations on the symmetry
structure of the wave function.  Numerical tests are presented and
possible generalizations are sketched, in particular for quantum dots,
i.e., particles interacting both among themselves and with an external
potential.

Let us consider first $N$ identical particles of mass $m$ whose
interaction does not depend on their spins.  This corresponds to the
Hamiltonian
\begin{equation}
H_{N}(m,g)=\sum_{i}{\vec{p}_{i}^2\over 2m}+g\sum_{i<j}V(r_{ij}),
\label{HN}
\end{equation}
where $r_{ij}=\vert \vec{r}_{j}-\vec{r}_{i}\vert$ is the distance 
between two particles. The case of particles with different masses is 
treated in Ref.~\cite{BAMA93}.

A first decomposition of this Hamiltonian is \cite{FIRU66,ADRI82}
\begin{equation}
H_{N}(m,g)= {1 \over N-2}
\sum_{i} H_{N-1}^{(i)}\left({N-1\over N-2}m,g\right), \label{HNH2}
\end{equation}
where the constituent mass in $H_{N-1}$ is increased, since the kinetic
energy $\vec{p}_{1}^2/(2m)$, for instance, is shared by $(N-1)$ terms.
The $i^{\rm th}$ particle is absent from $H_{N-1}^{(i)}$. 
Using the variational principle with the ground state of the $N$-body
system as trial wave-function leads to an inequality on the
ground-state energies $E_{N}$ \cite{FIRU66,ADRI82}.
\begin{equation}
E_{N}(m,g) 
\ge {N\over N- 2} E_{N-1}\left({N-1\over 
N-2}m,g\right),
\label{ENE2}
\end{equation}
which can be rewritten differently, using the obvious identity
$E(\alpha m,g)=E(m,\alpha g)/\alpha$.  This inequality is never
saturated, because the overall translation energy of the $(N-1)$-body
subsystems within the $N$-body system is neglected.  An improvement
consists of replacing the decomposition (\ref{HNH2}) by the identity
\begin{equation}
\widetilde{H}_{N}(m,g)={1\over
N-2}\sum_{i}\widetilde{H}_{N-1}^{(i)}\left({N\over 
N-1}m,g\right),
\label{HNtildeHN-1}
\end{equation}
relating the translation-invariant Hamiltonians
\begin{equation}
\widetilde{H}_{N}=H_{N}-{[\sum\vec{p}_{i}]^2\over 2 N m}\cdot
\label{HNtilde}
\end{equation}
This leads to the new inequality
\begin{equation}
E_{N}(m,g)
\ge  {N\over N-2} E_{N-1}\left({N\over N-1}m,g\right).
\label{ENtildeE2}
\end{equation}
As $Nm/(N-1)$ is smaller than $(N-1)m/(N-2)$ for $N\ge3$, the bound
(\ref{ENtildeE2}) is  better than (\ref{ENE2}), since
any binding energy in a given potential is a decreasing function of
the constituent mass. 

By recursion, an inequality such as (\ref{ENE2}) or (\ref{ENtildeE2})
provides a bound on $E_{N}$ in terms of the 2-body energy $E_{2}$.  In
the case of the harmonic oscillator for bosons, the bound derived from
(\ref{ENtildeE2}) is saturated, whereas the bound derived from
(\ref{ENE2}) is by a factor $\sqrt 2$ smaller than the exact result at
large $N$.

Some numerical illustration of the inequality (\ref{ENtildeE2}) for
bosons is provided in Table~\ref{table1}.  Other examples are given in
Ref.~\cite{HAPO67}.  

In the case of fermions, the bound (\ref{ENtildeE2}) is not
satisfactory.  Indeed, it is not necessarily the ground state of
$\widetilde{H}_{N-1}$ which is relevant, but the lowest state with
quantum numbers compatible with the symmetry of the $N$-body state of
interest. Thus looking at the structure of the wave function seems 
important to improve the inequalities.

Since our Hamiltonian acts in the orbital Hilbert space only, the
behavior of the wave function with respect to the space
variables has to be specified by an irreducible representation of the
symmetry group S$_N$. This corresponds to a partition
$[\lambda]=[\lambda_1,\lambda_2,\dots,\lambda_n]$ of the $N$ particles
defining an invariant orbital subspace
spanned by orthogonal states which are associated to a Young tableau
or a Yamanouchi symbol $r$ \cite{HA64}. Both specify the 
transformation under permutation, and the matrix element
of any transposition operator $P_{ij}$ is  known analytically. For
example, in the case of three particles, we can have the following
partitions 
\begin{equation}
    [3], \qquad [2,1], \qquad [1,1,1]. 
\end{equation}
The first and the third ones are in the familiar symmetric and
antisymmetric representations of dimension 1, respectively.  The second
one, of dimension 2, is the mixed symmetry representation spanned by
the two states $r=(2,1,1)$ or $r=(1,2,1)$ which are respectively
symmetric and antisymmetric in the exchange of particles 1 and 2.

 Each orbital state $\left\vert N,[\lambda],r
 \right\rangle_{\text{o}}$ must be associated with a similar state
 $\vert N,[\bar\lambda],\bar r \rangle_{\text{i}}$ for the intrinsic
 degrees of freedom such as spin, isospin, color, etc.  The coupling
 of $[\bar\lambda]$ to $[\lambda]$ gives a symmetric representation
 $[N]$ of S$_N$ for bosons or a fully antisymmetric representation
 $[1^N]$ for fermions.  This implies that $[\bar\lambda],\bar r$ are
 identical to $[\lambda],r$ when we are dealing with bosons, while for
 fermions, $[\bar\lambda],\bar r$ correspond to a Young tableau with
 rows and columns interchanged \cite{HA64}.
 
The S$_{N}$ Clebsch--Gordan coefficients
 $\langle [\lambda]r[\lambda]r\left\vert [N]\right\rangle$ and $\langle
 [\lambda]r[\bar\lambda]\bar r\left\vert [1^N]\right\rangle$ are 
 known explicitly \cite{HA64}, and the $N$-body state reads
 ($\epsilon=1$ for bosons, $-1$ for fermions)
\begin{equation}
\left\vert N,[\lambda]\right\rangle_{\epsilon} = d_{[\lambda]}^{-1/2} 
\sum_r \epsilon^{\sigma}\left\vert
N,[\lambda],r \right\rangle_{\text{o}}\left\vert
N,[\bar\lambda],\bar r \right\rangle_{\text{i}},
\label{fo}
\end{equation}
where
\begin{equation}
d_{[\lambda]} = N! \,{\prod_{1\le i< j\le n}
(\lambda_i-\lambda_j+j-i)
\over
\prod_{1\le i\le n} (\lambda_i+n-i)!}
\raise 2pt\hbox{,}
\end{equation}
is the dimension of the representation $[\lambda]$ and $\sigma$ is the
number of permutations that have to been performed to obtain the
Yamanouchi symbol from the normal one \cite{HA64}.  To obtain the
($N-1$)-body parts of state (\ref{fo}), one has now just to remove the
$N^{\text{th}}$ particle in each symbol $r$, thus leading to a
($N-1$)-Yamanouchi symbol that belongs to a different orbital symmetry
$[\lambda]_p = [\lambda_1, \dots,\lambda_{p-1},\,\mbox{$\lambda_p-1$},\,
\lambda_{p+1},\dots, \lambda_n]$, where $p$ is the row number of
particle $N$ in the original Young tableau.  Moreover, this new symbol
needs $a={\sum_{q>p=1}^N \lambda_q}$ less permutations than $r$ to
appear in the normal form.  Hence the $N$-body wave function
(\ref{fo}) can be rewritten as
\begin{equation}
\label{stateNvsN-1}    
\sum_p
{d_{[\lambda]_p}^{1/2} \over d_{[\lambda]}^{1/2}} \epsilon^a \left\vert
\left(N-1,[\lambda]_p\right)_{\epsilon} 
\left(1,[1]\right)  \right\rangle_{[\lambda],\epsilon},
\end{equation}
with $p$ running over all the lines of $[\lambda]$ where a box can be
dropped.  When this is used with the decomposition (\ref{HNtildeHN-1})
of the Hamiltonian, it leads to a new and very general inequality
\begin{equation}
E_N^{[\lambda]}(m,g) \ge {N\over N-2} \sum_p {d_{[\lambda]_p} \over
d_{[\lambda]}}E_{N-1}^{[\lambda]_p}\left({N\over N-1}m,g\right).
\label{Eours}
\end{equation}

If one considers, as a first application, a system of spinless bosons
for which the orbital Young pattern $[\lambda]$ is necessarily
symmetric, the inequality (\ref{Eours}) coincides with our previous 
result (\ref{ENtildeE2}). 

Consider now fermions with spin 1/2.  The conjugate partition $[\bar
\lambda]$ of $[\lambda]$ is an irreducible representation of U$(2)$
and contains at most 2 rows which are related to the spin $S$ by the
relations $ \bar \lambda_1 + \bar \lambda_2 = N$ and $\bar \lambda_1 -
\bar \lambda_2 = 2S$.  Writing our inequality (\ref{Eours}) with
$[\bar\lambda]$ gives the following result: the ground
state energy of $N$-spin 1/2 fermions with total spin $S$ is bound in
terms of binding energies of $(N-1)$-body systems with neighboring
spins $S\pm 1/2$
\begin{eqnarray}
    \label{EvsN-1F}
E_N^S&&(m,g)\ge {N-1\over N(N-2)(2S+1)}\times\nonumber\\
&&\left[ S(N+2S+2)
E_{N-1}^{S-1/2}\left(m,{Ng\over N-1}\right)\right. \\ 
&&\ + \left.(S+1)(N-2S)
E_{N-1}^{S+1/2}\left(m,{Ng\over N-1}\right)\right]\cdot\nonumber
\end{eqnarray}
The generalization is straightforward: the energetically favored state
has the most symmetric space partition $[\lambda]$ that admits a
conjugate partition $[\bar\lambda]$ which can be a representation of
U$(\Omega)$, where $\Omega$ is the number of intrinsic degrees of
freedom.  As a result, $[\bar\lambda]$ has a maximum of $\Omega$ rows
and $[\lambda]=[\Omega^\nu,N-\nu\Omega]$ for the $N$-body ground state
with $\nu$ the integer part of $N/\Omega$.  Denoting by $E_{N-1}^0$
the binding energy of ($N-1$) particles with the favored orbital
symmetry $[\Omega^\nu,N-\nu\Omega-1]$ and by $E_{N-1}^1$ the energy
associated to the first excited partition
$[\lambda]=[\Omega^{\nu-1},\Omega-1,N-\nu\Omega]$, the relation
(\ref{Eours}) takes the form
\begin{eqnarray}
E_N&&(m,g) \ge {N-1\over N(N-2)(1+\Omega+\nu\Omega-N)}\times\nonumber\\
&&\left[ (N-\nu\Omega)(1+\Omega+\nu+\nu\Omega-N)
E_{N-1}^0\left(m,{Ng\over N-1}\right)\right. \nonumber\\ 
&&+ \left.\nu(\Omega+1)(\Omega+\nu\Omega-N)
E_{N-1}^1\left(m,{Ng\over N-1}\right)\right]\cdot
\end{eqnarray}

We have calculated the lower bound and the exact energy with a
selection of potentials, using the method of Gaussian expansion
described for instance in Ref.~\cite{FLRI95}.  The variational
parameters are determined by a stochastic optimization \cite{VASU95}. 
The method has proved to be powerful and reliable.
  
In  Table~\ref{table1}, we show the ratio of the accurately
computed energy to the lower bound for monotonic short-range
potentials.  For $N=3$ and $S=1/2$, each pair is in an equal-weight
admixture of singlet and triplet, while $S=3/2$ involves only triplet
states.  For $N=4$ and $S=0\,(2)$, each 3-body subsystem has
$S=1/2\,(3/2)$. 

It appears from Table~\ref{table1} that the bound is close to the 
exact result, especially for antisymmetric orbital wave functions 
($N=3$, $S=3/2$ or $N=4$, $S=2$) and for deeply bound states.

\begin{figure}
    \epsfbox{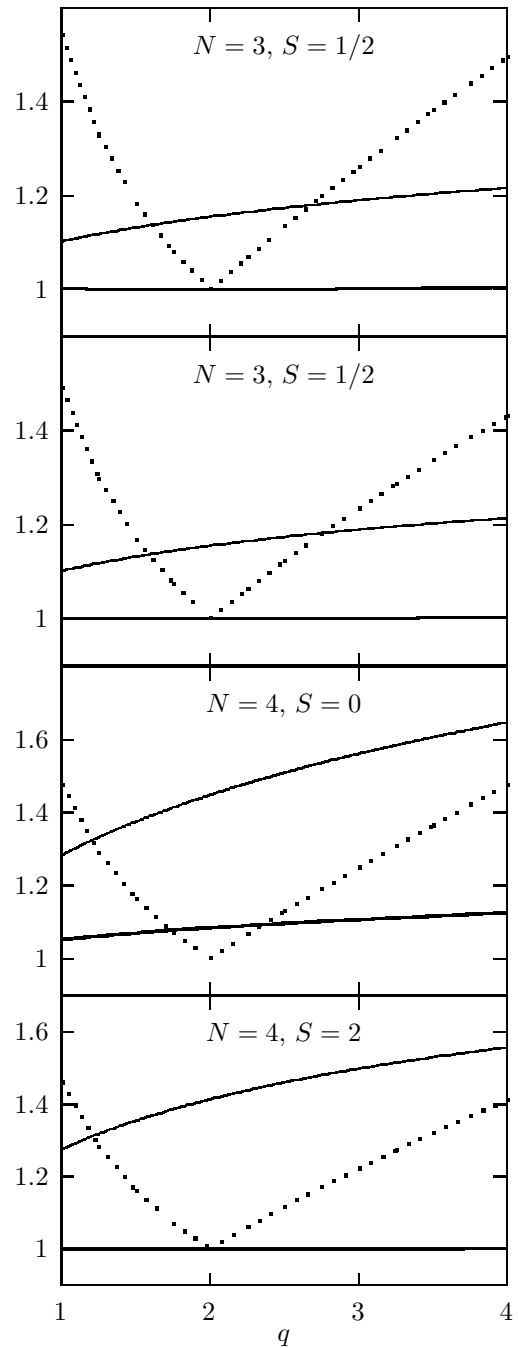}
	 \caption{Comparison of various lower
	 bounds for $N=3$ or 4 fermions with total spin $S$ , in the
	 case of power-law potentials $r^q$.  The quantity shown is
	 the ratio of the exact energy to our bound (thick line), to
	 that of Basdevant and Martin (dotted line) and that of
	 L\'evy-Leblond (thin line).}
\label{figfermion} 	
\end{figure}

The results in Fig.~\ref{figfermion} correspond to power-law 
potentials $r_{ij}^q$ with $q\ge1$. For comparison, we also display 
the bound by L\'evy-Leblond \cite{FIRU66}, initially designed for 
large systems but also applicable at small $N$. The decomposition
\begin{equation}
H_{N}(m,g)
   ={1\over 2}\sum_{i}\sum _{j\neq i}
 \left[ {\vec{p}_{j}^2\over (N-1) m} + g V(r_{ji})\right]
  \label{HNLL}
\end{equation}
expresses $H_{N}$ 
in terms of Hamiltonians with $(N-1)$ independent particles. For fermions, 
one gets
\begin{equation}
E_{N}(m,g)
   \ge {N\over 2}f_{N-1}(m(N-1),g),
  \label{HNferm}
\end{equation}
where $f_{N}$ is the cumulated energy of a system of $N$ independent
fermions, a notation borrowed from Ref.~\cite{BAMA96}. In 
Eq.~(\ref{HNferm}), the translation energy of the $(N-1)$-body 
subsystem is neglected. 

Also shown in Fig.~\ref{figfermion} is the bound of Basdevant and Martin 
\cite{BAMA96}, who started from the identity
\begin{equation}
    \label{HO-identity}
 \sum_{i}\vec{p}_{i}^2 +  \vec{r}_{i}^2=
 {\vec{P}^2\over N} +N\vec{R}^2 + T_{\text{r}}+
 {1\over N}\sum_{i<j} r_{ij}^2,
\end{equation}
where $\vec{P}$ is the total momentum, $\vec{R}$ the center-of-mass
position and $ T_{\text{r}}$ the relative kinetic energy, and
succeeded to generalize it in the form of inequalities when the power
$q=2$ is replaced by another power $q\ge 1$.  Then the Hamiltonian
(\ref{HN}) or its translation-invariant part (\ref{HNtilde}) with
$V\propto r_{ij}^q$ is bounded (on both sides) by independent-particle
Hamiltonians with a potential proportional to $r_{i}^q$.  This leads
to
\begin{eqnarray}
&& E_N\ge 2^{{ q-4\over  q+2}}
\left[N^{{ 2\over  q+2}} 
f_{N}-N^{{ 4-q\over  q+2}} E_{2}\right]\nonumber\\
&&E_N\le2^{{ -q\over  q+2}}\left[
N^{{ 2\over q+2}} f_{N}-N^{{ q\over 
 q+2}}E_{2}\right],
\label{inegBM}
\end{eqnarray}
for $1\le q\le 2$. The inequalities are reversed for $q\ge 2$. 

Clearly from Fig.~\ref{figfermion}, Basdevant and Martin always give 
the best bound near $q=2$, but their result quickly deteriorates when 
the potential departs from the harmonic case. Our inequality is saturated at 
$q=2$ for $N=3$ fermions and for some spin configurations of $N=4$, 
but saturation is lost at larger $N$. For small systems, our bound 
dramatically improves that of L\'evy-Leblond. 

Our result can be extended to more general Hamiltonians, for instance 
with spin--spin interaction or with external constraints. More 
details will be given elsewhere. Let us just mention that our bound 
can be applied to the  ``quantum dot'' systems
\begin{equation}
H=\sum_{i=1}^N\,{\vec{p}_{i}^2\over2m}+{m\omega^2\over 2} \vec{r}_{i}^2
+\sum_{i<j}{e^2\over r_{ij}}\raise 2pt\hbox{,}
\end{equation}
once the center-of-mass contribution (a mere harmonic oscillator) is 
removed. In the case of $N=3$ electrons with mass $m=1$ and charge $e=1$, one 
obtains, for $S=3/2$ and orbital momentum and parity $L^P=1^+$, a bound
$\epsilon\ge 0.1668$ if the oscillator frequency is $\omega=0.01$ and 
$\epsilon\ge 54.968$ if $\omega=10$, to be compared to the exact values
$\epsilon=0.1680$ and $\epsilon=54.973$.
For $N=4$, $S=2$, $L^P=0^-$, one obtains $\epsilon\ge 0.297$ if 
$\omega=0.01$  and $\epsilon\ge 84.895$ if $\omega=10$, close to the exact 
 values $\epsilon=0.299$ and $\epsilon=84.907$. The bound is better 
 for large $\omega$, as the system becomes a pure oscillator.

To conclude, we have derived a lower bound on the energy of
$N$-fermion systems, which is independent of the shape of the
potential. For small systems, it improves significantly a previous
bound expressed in terms of $(N-1)$ independent fermions.

Saturation is not always obtained for the harmonic 
oscillator. When our bound is iterated to express the $N$-body energy 
in terms of the 2-body energy, the correct large $N$ behavior is not 
reached. Thus there is still much room for improvement.

{\bf Acknowledgements}. We would like to thank Ica Stancu, Jean-Louis 
Basdevant, Andr\'e Martin and Piet van Isacker for very useful comments.

%%%%%%%%%%%%%%

%%%%%%%%%%%%%%%%%%%%%%%%%%%%%
\begin{table}[htbp]
\begin{tabular}{cccccccc}%
&$g$&\ 3, B, 0\ \ &\ 3, F, 1/2\ &\ 3, F, 3/2&
\ 4, B, 0\ \ &\ 4, F, 0\ &\ 4, F, 2\ \cr
\hline
Y & 8  & 0.933 & 0.673 & 0.759 & 0.966 & 0.743 & 0.855\cr
  & 15 & 0.943 & 0.757 & 0.930 & 0.971 & 0.806 & 0.964\cr
\hline
G & 10 & 0.996 & 0.960 & 0.887 & 0.998 & 0.792 &0.945\cr
  & 20 & 0.999 & 0.995 & 0.994 & 0.999 & 0.898 &0.997\cr
\hline
E & 6  & 0.988 & 0.906 &0.843  & 0.994 & 0.795 &0.913 \cr
  & 12 & 0.994 & 0.974 &0.982  & 0.997 & 0.886 &0.991 \cr
\end{tabular}
\vspace{0.2cm}
\caption{Results for Yukawa (Y), Gaussian (G) and exponential
      (E) potentials.  The range parameter is set to unity by
      rescaling.  The quantity shown is the ratio of the computed
      energy to the bound.  In the first line, the entries are: the
      number of particles, the boson (B) or fermion (F) character, and
      the total spin.  The 4-body energy is compared to the 3-body
      one, as per Eq.~(\protect\ref{ENE2}) for bosons and
      (\protect\ref{EvsN-1F}) for fermions.  The strength $g$ refers
      to the Hamiltonian for $N=3$ particles.}
\label{table1}
\end{table}

%%%%%%%%%%%%%%%%%%%%%%%%%%%%%%
%%%%%%%%%%%%%%%%%%%%%%%%%%%%

\end{document}